\documentclass[twocolumn,showpacs,pra,superscriptaddress,amsmath,amssymb]{revtex4}

\usepackage{booktabs}
\usepackage{color}
\usepackage{graphicx}% Include figure files
\usepackage{dcolumn}% Align table columns on decimal point
\usepackage{bm}% bold math
\begin{document}

\title{Topological Fulde-Ferrell Superfluids in Triangular Lattices}

\author{Long-Fei Guo}
\affiliation{College of Physical Science and Technology, Sichuan University, 610064,
Chengdu, China}
\affiliation{Key Laboratory of High Energy Density Physics and Technology of Ministry of
Education, Sichuan University, Chengdu, 610064, China}

\author{Peng Li}
\email{lipeng@scu.edu.cn}
\affiliation{College of Physical Science and Technology, Sichuan University, 610064,
Chengdu, China}
\affiliation{Key Laboratory of High Energy Density Physics and Technology of Ministry of
Education, Sichuan University, Chengdu, 610064, China}

\author{Su Yi}
\email{syi@itp.ac.cn}
\affiliation{CAS Key Laboratory of Theoretical Physics, Institute of Theoretical Physics,
Chinese Academy of Sciences, P.O. Box 2735, Beijing 100190, China}
\affiliation{School of Physics, University of Chinese Academy of Sciences, P. O. Box 4588, Beijing 100049, China}

\date{\today}

\begin{abstract}
Fulde-Ferrell (FF) Larkin-Ovchinnikov (LO) phases were proposed for
superconductors or superfluids in strong magnetic field.
With the experimental progresses in ultracold atomic systems,
topological FFLO phases has also been put forward, since it is a natural
consequence of realizable spin-orbital coupling (SOC).
In this work, we theoretically investigate a triangular lattice
model with SOC and in-plane field. By constructing the phase diagram,
we show that it can produce topological FF states with Chern numbers,
$C=\pm1$ and $C=-2$. We get the phase boundaries by the change of the sign of Pfaffian.
The chiral edge states for different topological FF phases are also elucidated.

\end{abstract}


\maketitle

%\tableofcontents

\section{Introduction\label{sec:1}}

Cooper pair was first proposed in 1956 to explain superconductors \cite{Cooper}.
It describes a pair of fermions bound together due to attractive interaction.
The fermions have opposite momentum so that the pair has zero momentum totally.
However, Cooper pair with finite center-of-mass momentum may also exist
in the presence of strong magnetic field. This consideration led to an exotic
superconductor with inhomogeneous order parameter in real space, known as
Fulde-Ferrell \cite{Fulde} and Larkin-Ovchinnikov (FFLO) phases \cite{Larkin}. There are two types
of FFLO phases: phase modulated FF state and spatial modulated LO state. In the past two decades, FFLO phases attract tremendous interests in both experiment and theory \cite{Casalbuoni, Kenzelmann, Liao, Li, Hu1, Radzihovsky}. But only ambiguous experiment evidences from heavy-fermion superconductors and organic superconductors are available \cite{Matsuda, Shimahara, Koutroulakis}.

On the other hand, topology is also a hot topic in condensed matter field
for several decades \cite{Xiao, Hasan, Qi}. Recently, the spin-orbit coupling (SOC)
has been realized by ultracold atoms as condensates or on an optical lattice \cite{Lin, Wang, Cheuk, Wu},
which paves the way to the topological FFLO states \cite{Liu, Qu, Zhang, Chen, Cao, Hu2}.
Theoretical researches show that topological FFLO states can be induced in one and two dimensional
Fermi gas. And according to the bulk-edge correspondence,
edge modes are supported when there are boundaries \cite{Hatsugai, Bernvig}.

In two dimensions, topological FF state with Chern number $C=1$ can be produced for cold atoms in a square lattice \cite{Qu, Chen}. To realize topological state with higher Chern number, one can resort to complicated hoppings \cite{Sarma} or lattices \cite{Huang}. Nonetheless, the simple triangular lattice favors some topologically nontrivial states, it can produce topological state with higher Chern number by merely the nearest-neighbor hoppings \cite{Kai2, Iskin}. In this work, we investigate a system with SOC and in-plane Zeeman field on the triangular lattice to achieve topological FF states. We found, in the noninteracting case, our system with nearest-neighbor hoppings supports the gapped Chern insulators with Chern number $C=1$ as well as $C=-2$. The boundaries of both of the topological phases are ellipses in a two-parameter plane constituted by the in-plane and out-plane fields. In the presence of on-site attractive s-wave interactions, we solve the system by self-consistent equations. The non-uniform FF superfluid states are obtained in a large area. And we confirm that all the FF states are topologically nontrivial. Their Chern numbers are $C=1,-1$ or $-2$. We use a set of signs of Pfaffians at high symmetry points in the first Brillouin zone (1st BZ) to characterize the different topological phases. Each one of these signs of Pfaffians changes with the energy gap closing and reopening at the corresponding point. We calculate the chiral edge states, whose wave functions are spatially localized at the edges in open boundary situation, which can help us to confirm that the bulk is in a topologically nontrivial state. Different FF phases exhibit different pairs of chiral edge states. The edge current is directly determined by Chern number, i.e. the summation of chiralities of the edge modes.

This paper is organized as follows. In Sec. \ref{sec:2}, we introduce a system in the triangular lattice with SOC and in-plane field and solve it by self-consistent method. In Sec. \ref{sec:3}, we construct the phase diagram in the noninteracting and interacting case. The topological FF phases with $C=-1,1$ and $-2$ are obtained in the presence of interaction. The chiral edge states in distinct topological FF phases are also illustrated. At last, we give a brief summary.

\section{Model Hamiltonian} \label{sec:2}

\begin{figure}
\includegraphics[width=0.47\textwidth]{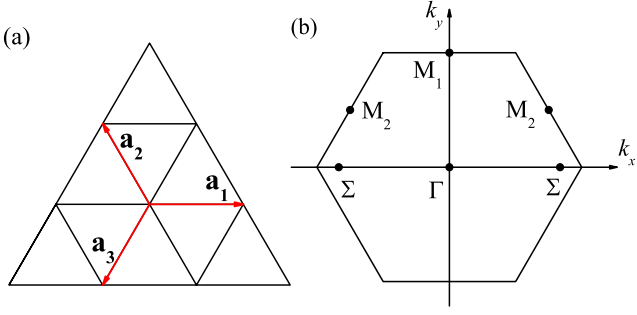}
\caption{\label{fig:1}(Color line) (a) Triangular lattice with direction vectors $\mathbf{a}_{1}$, $\mathbf{a}_{2}$, and $\mathbf{a}_{3}=-(\mathbf{a}_{1}+\mathbf{a}_{2})$. (b) Brillouin zone of triangular lattice.
}
\end{figure}

We consider a two-dimensional (2D) spin-orbit coupled fermionic gas trapped in a triangular lattice subjected to an external magnetic field. In the position space, the model Hamiltonian reads
\begin{equation}
H=\sum_{\langle i,j\rangle}\phi _{i}^{\dagger }H_{s}\phi
_{j}+H_{\text{int}},
\end{equation}
where $\phi _{i}^{\dagger }=\left( c_{i,\uparrow },c_{i,\downarrow }\right) $ with $c_{i,\sigma }^{\dagger }(c_{i,\sigma })$ being the creation (annihilation) operator for the spin $\sigma $ particle at site $i$, $H_{s}$ and $H_{\rm int}$ represent the single-particle and interaction Hamiltonian, respectively, and the summation is over the nearest-neighbor pairs. More specifically, we assume that the single-particle Hamiltonian takes the form
\begin{equation}
H_{s}=t^{i,j}\sigma _{z}+i t_{\rm so}^{i,j}(\mathbf{v}_{i,j}\times \boldsymbol{\sigma})_{z}
+(\mathbf{h}\cdot\boldsymbol{\sigma}-\mu)\delta_{i,j},  \label{equ:2}
\end{equation}
where $\sigma_{\alpha}$ ($\alpha=x,y,z$) are a Pauli matrices, $t^{i,j}$ is the hopping matrix elements which, when combined with $\sigma$, gives rise to opposite signs for the spin-up and -down particles \cite{Xiongjun}, $t_{\rm so}^{i,j}$ are the Rashba SOC coefficients, $\mathbf{v}_{i,j}$ are the vectors connecting lattice sites $i$ and $j$, $\mathbf{h}=(h_{x},0,h_{z})$ is the magnetic field which includes an in-plane component $h_{x}$ along the $x$ direction and an out-of-plane one $h_{z}$, and $\mu$ is chemical potential. The $s$-wave interaction Hamiltonian can be expressed as
$$H_{\text{int}}=-U\sum_{i}c_{i,\uparrow }^{\dagger }c_{i,\uparrow }c_{i,\downarrow
}^{\dagger }c_{i,\downarrow },$$
where $U>0$ represents attractive interaction. Finally, in the triangular lattice, the lattice basis vectors are defined as $\mathbf{a}_{1}=(1,0)$, $\mathbf{a}_{2}=(-1/2,\sqrt{3}/2)$, and $\mathbf{a}_{3}=(1/2,\sqrt{3}/2)$ which, as shown in Fig.~\ref{fig:1}(a), represent the displacements from one site to the nearest sites. Correspondingly, we obtain a hexagonal first Brillouin zone (1BZ) with reciprocal lattice basis vectors $\mathbf{G}_{1}=(0,4\pi /\sqrt{3})$ and $\mathbf{G}_{2}=(2\pi ,2\pi \sqrt{3})$.

The attractive interaction may lead to the Bardeen-Cooper-Schrieffer (BCS) superfluid. Particularly, in the presence of the in-plane Zeeman field along the $x$ direction, the Fermi surface becomes asymmetric along the $y$ axis. Consequently, the BCS pairs may carry a finite center-of-mass momentum $Q_{y}$ along the $y$ direction~\cite{Zheng}. Such state is described by the FF order parameter, which, in the position space, is defined as
$$\mathbf{\Delta }_{i}=U\left\langle c_{i,\downarrow }c_{i,\uparrow }\right\rangle =\Delta e^{i\mathbf{Q}\cdot
\mathbf{R}_{i}}$$
with $\mathbf{Q}=(0,Q_{y})$~\cite{momentqy}. After transformed into the momentum space, the mean-field Hamiltonian in the Nambu representation reads
$$H=\frac{1}{2}\sum_{\mathbf{k}}\Psi _{\mathbf{k}}^{\dagger
}H_{\text{BdG}}\Psi _{\mathbf{k}}+N\left(\frac{\left\vert \mathbf{\Delta }_{i}\right\vert
^{2}}{U}-\mu \right),$$
where $\Psi _{\mathbf{k}}=\left( c_{\mathbf{k}+\mathbf{Q}/2,\uparrow
},c_{\mathbf{k}+\mathbf{Q}/2,\downarrow },-c_{-\mathbf{k}+\mathbf{Q}
/2,\downarrow }^{\dagger },c_{-\mathbf{k}+\mathbf{Q}/2,\uparrow }^{\dagger
}\right) ^{T}$ is the Nambu spinor with ${\mathbf k}=(k_{x},k_{y})$, $N$ is the total number of lattice sites, and the Bogoliubov-de Gennes (BdG) Hamiltonian is
\begin{align}
H_{\text{BdG}}=&\,(a_{\mathbf{k}}+h_{z})\tau_{0}\otimes\sigma_{z}+b_{\mathbf{k}}\tau_{z}
\otimes\sigma_{z}+c_{\mathbf{k}}\tau_{z}\otimes\sigma_{x}\nonumber\\
&\,+(d_{\mathbf{k}}+h_{x})\tau_{0}\otimes\sigma_{x}+e_{\mathbf{k}}\tau_{z}
\otimes\sigma_{y}+f_{\mathbf{k}}\tau_{0}\otimes\sigma_{y}\nonumber\\
&\,+\Delta\tau_{x}\otimes\sigma_{0}-\mu\tau_{z}\otimes\sigma_{0}\label{equ:3}
\end{align}
with $\otimes$ being the Kronecker product and $\tau_{\alpha}$ ($\alpha=x,y,z$) being the Pauli matrices acting on the particle-hole space. The explicit expressions for the elements of $H_{\text{BdG}}$ matrix, $a_{\mathbf k}$, $b_{\mathbf k}$, $c_{\mathbf k}$, $e_{\mathbf k}$, $f_{\mathbf k}$, and $g_{\mathbf k}$, can be found in the Appendix \ref{appa}. We note that the BdG Hamiltonian only possesses a particle-hole symmetry, $\Xi H_{\text{BdG}}(\mathbf{k})\Xi ^{-1}=\Lambda H_{\text{BdG}}^{\ast }(\mathbf{k})\Lambda
=-H_{\text{BdG}}(-\mathbf{k})$, where $\Xi=\Lambda K$ with $\Lambda=i\sigma_{y}\otimes\tau_{y}$ and $K$ being the complex conjugation operator. It can be verified that $\Xi^{2}=1$. Apparently, the system belongs to the class D according to Wigner-Dyson symmetry classification of random matrix~\cite{Altland} which has a topological invariant Z in two dimension~\cite{Schnyder, Ryu, Kiteav1}.

Following the standard treatment, we diagonalize the BdG Hamiltonian, $H_{\rm BdG}|\psi_{\alpha}^{\nu}({\mathbf k})\rangle=E_{\alpha,{\mathbf k}}^{\nu}|\psi_{\alpha}^{\nu}({\mathbf k})\rangle$, which leads to the   quasiparticle eigenenergy $E_{\alpha,{\mathbf k}}^{\nu}$ and the quasiparticle wave function $|\psi_{\alpha}^{\nu}({\mathbf k})\rangle$, here $\nu=\pm$ represent the particle ($+$) and hole ($-$) bands, $\alpha=1$ and $2$ denote the upper ($1$) and lower ($2$) helicity branches. Now, the thermodynamic potential at temperature $T$ can be calculated through
\begin{align}
\Omega =&\;\frac{N}{U}\left\vert \Delta \right\vert ^{2}-N\mu+\frac{1}{2}\sum_{\alpha,\mathbf{k}}E_{\alpha,{\mathbf k}}^{-}  \nonumber \\
&\;-k_{B}T \sum_{\alpha,\mathbf{k}}\ln \left(1+e^{E_{\alpha,{\mathbf k}}^{-}/(k_{B}T)}\right)\text{ ,}
\label{equ:4}
\end{align}
where $k_{B}$ the Boltzmann constant and the summation is restricted to the hole bands ($\nu=-$) due to the inherent particle-hole symmetry in the Nambu spinor representation. The order parameter can be numerically determined by the mean-field saddle equations $\partial\Omega /\partial \Delta =0$, $\partial \Omega /\partial Q_{y}=0$, as well as the equation for the conservation of the total particle number $\partial\Omega/\partial\mu=-N$. Finally, the topological property of the system is characterized by the Chern number \cite{Thouless, Xiao}
\begin{equation}
C=\sum_{\alpha}\frac{1}{2\pi }\int dk_{x}dk_{y}\mathbf{z}\cdot {\nabla}_{\mathbf{k}}\times \mathbf{A}_{\alpha}^{-}(\mathbf{k})\text{,}\label{equ:5}
\end{equation}
where $\mathbf{A}_{\alpha}^{-}(\mathbf{k})=i\left\langle \psi_{\alpha}^{-}(\mathbf{k}%
)\right\vert {\nabla}_{\mathbf{k}}\left\vert \psi_{\alpha}^{-}(\mathbf{k}%
)\right\rangle $ is the vector potential.

\section{Results} \label{sec:3}
In this section, we present the results about the zero-temperature quantum phases of our model. For convenience, we assume that the hopping matrix element is site independent, i.e., $t=t^{i,j}$ and $t_{\rm so}=t_{\rm so}^{i,j}$. Moreover, we select $t$ as the energy unit such that the model Hamiltonian is completely specified by the parameters $t_{\rm so}/t$, $h_{x}/t$, $h_{z}/t$, $U/t$, and $\mu/t$. In below, the value of the chemical potential is fixed at $\mu/t=1$ for simplicity.

\begin{figure}
\includegraphics[width=0.45\textwidth]{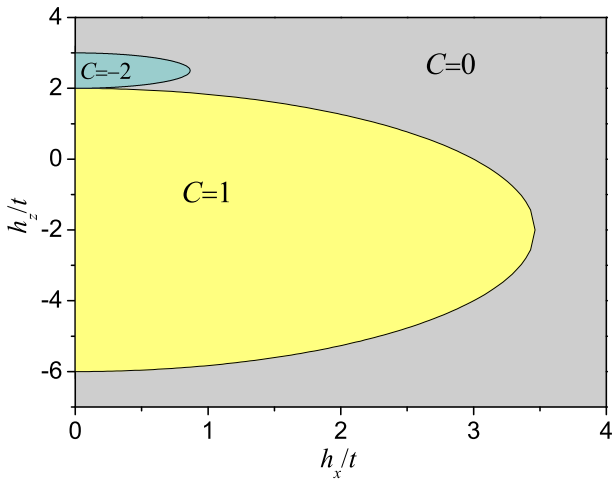}
\caption{\label{fig:2}(Color line) Quantum phases of the single-particle Hamiltonian Eq. (\ref{h1par}) with $t_{\rm so}/t=1$. Chern insulators are located within two ellipses as expressed by Eq. (\ref{ellipse1}) and (\ref{ellipse2}).}
\end{figure}

To start, let us briefly discuss the quantum phases of our model in the absence of interaction. In the momentum space, the single-particle Hamiltonian reduces to
\begin{eqnarray}
H_{s}=\sum_{\mathbf{k}}(d_{x}\sigma_{x}+d_{y}\sigma_{y}+d_{z}\sigma_{z}),\label{h1par}
\end{eqnarray}
where $d_{x}=\sqrt{3}t_{\rm so}(\sin{k_{3}-\sin{k_{2}}})+h_{x}$, $d_{y}=t_{\rm so}(2\sin{k_{1}}-\sin{k_{2}}-\sin{k_{3}})$, and $d_{z}=t(\cos{k_{1}}+\cos{k_{2}}+\cos{k_{3}})+h_{z}$ with $k_{\alpha}=\mathbf{k}\cdot\mathbf{a}_{\alpha}$. Due to the breaking of the time-reversal and chirality symmetries, the single-particle Hamiltonian belongs to the symmetry class C. Therefore, the possible topologically nontrivial ground state in 2D system is characterized by Z invariant~\cite{Schnyder,Ryu,Kiteav1}. To identify the topological state, we focus on the gapless points defined by
\begin{eqnarray}
|{\mathbf d}({\mathbf k})|=0,\label{gapless}
\end{eqnarray}
where ${\mathbf d}\equiv (d_{x},d_{y},d_{z})$. It can be shown that Eq.~(\ref{gapless}) gives rise to two ellipses,
\begin{align}
\frac{h_{x}^{2}}{3t_{\rm so}^{2}}+\left(\frac{h_{z}}{t}-\frac{5}{2}\right)^{2}=\frac{1}{4}
\label{ellipse1}
\end{align}
and
\begin{align}
\frac{4h_{x}^{2}}{3t_{\rm so}^{2}}+\left(\frac{h_{z}}{t}+2\right)^{2}=16,
\label{ellipse2}
\end{align}
which divide the $h_{x}h_{z}$ parameter plane into three regions. The topological property of each region is determined by the Chern number $C=\frac{1}{4\pi}\int_{\rm 1BZ}dk_{x}dk_{y}\mathbf{n}\cdot\partial_{k_{x}}\mathbf{n}\times\partial_{k_{y}}\mathbf{n}$, where $\mathbf{n}=\mathbf{d}/|\mathbf{d}|$ is a unit vector.

\begin{figure}
\includegraphics[width=0.47\textwidth]{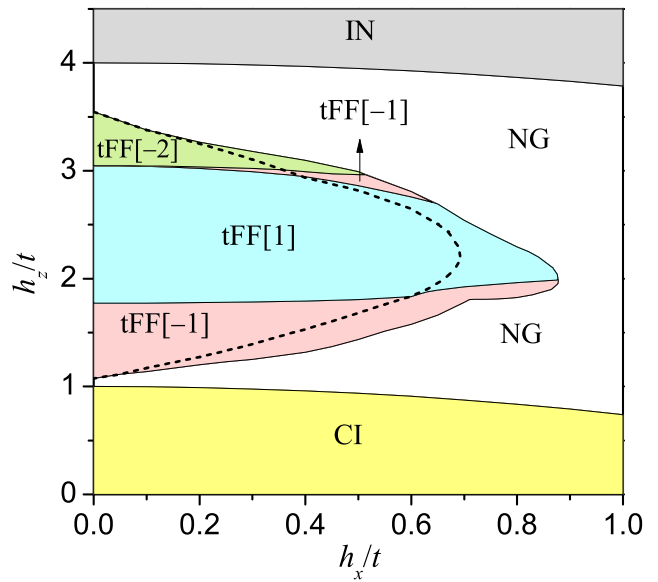}
\caption{\label{fig:3} (Color line) Zero-temperature phase diagram with $t_{\rm so}/t=1$, $U/t=5$, and $\mu/t=1$. The dashed curve represents the boundary between the gapped (left side) and gapless (right side) topological superfluid phases. The integers in the square brackets specify Chern numbers.}
\end{figure}

\begin{table}
\caption{Calculation of Chern number in different parameter region for the single particle system $H_{s}$.
\label{table1}}
\begin{ruledtabular}
\begin{tabular}{lcccccc}
parameter region & \multicolumn{4}{c}{$o(\mathbf{k}_{i})p(\mathbf{k}_{i})$} & $C$\\ \cline{2-5}
$ $& $\mathbf{k}_{1}$ & $\mathbf{k}_{2}$& $\mathbf{k}_{3}^{\pm}$ & $\mathbf{k}_{4}^{\pm}$ & $ $\\
\colrule
upper ellipse & 1 & -1 & -1 & -1 & -2\\
lower ellipse  & 1 & 1 & -1 & 1 & 1\\
%$h_{z}<-2-\sqrt{16-\eta^{2}}$ & -1 & 1 & -1 & 1 & 0\\
%$-2-\sqrt{16-\eta^{2}}<h_{z}<(5-\sqrt{1-\eta^{2}})/2$ & 1 & -1 & -1 & 1 & 0\\
%$h_{z}>1/2(5-\sqrt{1-\eta^{2}})$ & 1 & -1 & 1 & -1 & 0\\
\end{tabular}
\end{ruledtabular}
\end{table}

In Fig.~\ref{fig:2}, we plot the phase diagram of the Hamiltonian~(\ref{h1par}) for $t_{\rm so}/t=1$. Without loss of generality, only the result for $h_{x}>0$ is shown. As can be seen, the parameter regions enclosed by two ellipses are topologically nontrivial. Particularly, the Chern number inside the upper ellipse is $-2$. These topologically nontrivial states can be understood by noting that the Chern number can be reexpressed as~\cite{Tretiakov, Clement}
\begin{eqnarray}
C=\frac{1}{2}\sum_{\mathbf{k}_{i}}o(\mathbf{k}_{i})p(\mathbf{k}_{i}),
\end{eqnarray}
where $\mathbf{k}_{i}$ are the roots of the equations $d_{x}(\mathbf{k})=d_{y}(\mathbf{k})=0$ (see Appendix \ref{appb}), $o(\mathbf{k}_{i})$ is the chirality of the vector field $(d_{x}(\mathbf{k}),d_{y}(\mathbf{k}))$ around $\mathbf{k}_{i}$, and $p(\mathbf{k}_{i})=\text{sgn}(d_{z}(\mathbf{k}_{i}))$ is the polarity at $\mathbf{k}_{i}$. In Tab.~\ref{table1}, we summarize the contribution of each root to the Chern number.

\begin{figure}
\includegraphics[width=0.47\textwidth]{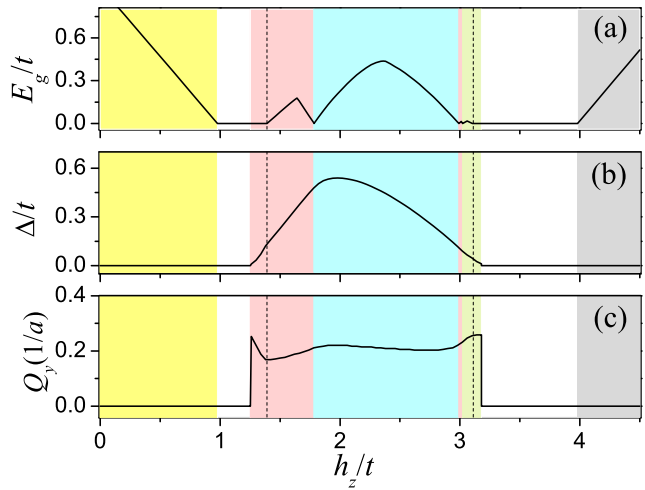}
\caption{\label{fig:4} (Color line) $h_{z}$ dependence of the quasiparticle band gap $E_{g}$ (a), superfluid order parameter $\Delta$ (b), and pairing momentum $Q_{y}$ (c) for $h_{x}/t=0.3$. Other parameters are the same as those in Fig.~\ref{fig:3}.}
\end{figure}

We now turn to study the superfluid phases of the system by taking into account the attractive $s$-wave interaction. Fig.~\ref{fig:3} summarizes the quantum phases in the $h_{x}h_{z}$ parameter plane with $U/t=5$ and $\mu/t=1$. Here the quantum phases are characterized by the superfluid order parameter $\Delta$, the center-of-mass momentum $Q_{y}$, the Chern number $C$, and the chemical potential $\mu$. Specifically, when $\Delta=0$, we may either have an insulating (IN) phase or a normal gas (NG) phase. For the former state, the chemical potential locates in the band gap; while, for the latter state, it lies in the band such that the excitation is gapless~\cite{Xu}. Moreover, an IN phase is Chern insulator (CI) if it is topologically nontrivial ($C=1$ here). Next, a superfluid state ($\Delta\neq 0$) with nonzero $Q_{y}$ is denoted by `FF'. In our model, the appearance of FF states is due to the in-plane magnetic field $h_{x}$ that deforms the Fermi surface~\cite{Zheng}. In fact, the standard Bardeen-Cooper-Schrieffer superfluid ($Q_{y}=0$) only exists at $h_{x}=0$. Additionally, all superfluid phases are found to be topologically nontrivial and they are further specified by the letter `t' and by Chern numbers in the square bracket. Finally, the dashed line in Fig.~\ref{fig:3} marks the boundary between gapped and gapless superfluid phases. More specifically, for small $h_{x}$, the energy of the lower helicity particle branch $E_{2,{\mathbf k}}^{+}$ is always positive for the gapped superfluid phases. However, $E_{2,{\mathbf k}}^{+}$ may become less than zero as $h_{x}$ is increased, which leads to the gapless superfluid phases \cite{Fan, Deng}.

In Fig.~\ref{fig:4}, we plot the $h_{z}$ dependence of the quasiparticle band gap $E_{g}\equiv{\rm max}\{0,{\rm min}(E_{2,{\mathbf k}}^{+})\}$, the superfluid order parameter $\Delta$, and the pairing momentum $Q_{y}$ for a fixed $h_{x}/t=0.3$. As can be seen, for small $h_{z}$, the Fermi surface locates in the band gap such that the system remains in the insulating phase. Then as $h_{z}$ is increased, $E_{2,{\mathbf k}}^{+}$ moves downward, while $E_{2,{\mathbf k}}^{-}$ moves oppositely, which leads to a vanishing $E_{g}$, signaling the entering of the NG phase. As one further increases $h_{z}$, the tFF phases emerge. Within the superfluid phases, the energy gap closes and reopens whenever a topological phase transition is encountered. Eventually, at very large $h_{z}$, the  large population difference between the spin-$\uparrow$ and -$\downarrow$ particles prohibits the formation of the superfluid pairs. As a result, the system falls into insulting phases again.

The phase boundaries between topological superfluid phases in Fig.~\ref{fig:3} can be alternatively determined by considering the high symmetry points $\mathbf{k}'$ that satisfy $\Xi H_{\text{BdG}}(\mathbf{k}')\Xi^{-1}=-H_{\text{BdG}}(\mathbf{k}')$. To this end, we introduce an auxiliary matrix $W(\mathbf{k}')\equiv H_{\text{BdG}}(\mathbf{k}')\Lambda$ \cite{Kiteav2, Ghosh} that is antisymmetric, i.e., $W^{T}({\mathbf k}')=- W({\mathbf k}')$. A topological index can then be defined as
\begin{equation}
P(\mathbf{k}')={\rm sgn}[{\rm Pf}[W(\mathbf{k}')]],\label{equ:8}
\end{equation}
where ${\rm Pf}[W(\mathbf{k}')]$ denotes the Pfaffian of $W(\mathbf{k}')$. Since ${\rm Pf}[W(\mathbf{k}')]=\pm\sqrt{\det[H_{\text{BdG}}(\mathbf{k}')]}$, $P(\mathbf{k}')$ will never change its sign unless the energy gap at $\mathbf{k}'$ is closed, indicating that $P(\mathbf{k}')$ is indeed topologically protected. This also suggests that the phase boundaries are determined by the condition $P({\mathbf k}')=0$ or, formally,
\begin{equation}
[d(\mathbf{k}')+h_{x}]^{2}+[a(\mathbf{k}')+h_{z}]^{2}-\Delta(h_{x},h_{z})^{2}=\mu ^{2}-f(\mathbf{k}')^{2}.\label{phaseb}
\end{equation}%
In our model, the high symmetry points include $\mathbf{\Gamma} =(0,0)$, $\mathbf{M}_{1}=(0,2\pi /\sqrt{3})$, $\mathbf{M}_{2}=(\pm \pi ,\pi /\sqrt{3})$, and $\mathbf{\Sigma}=(\pm2\cos^{-1} [-\cos(\sqrt{3}Q_{y}/4)/2],0)$, as shown in Fig.~\ref{fig:1}(a). Here $P(\mathbf{k}'=\Gamma )=-1$ remains unchanged and is irrelevant to the topological phase transitions. However, it can be verified that other high symmetry points $\mathbf{M}_{1}$, $\mathbf{M}_{2}$, and $\mathbf{\Sigma}$ leads to the three phase boundaries between the topological superfluid phases through Eq.~(\ref{phaseb}).

\begin{figure}
\includegraphics[width=0.47\textwidth]{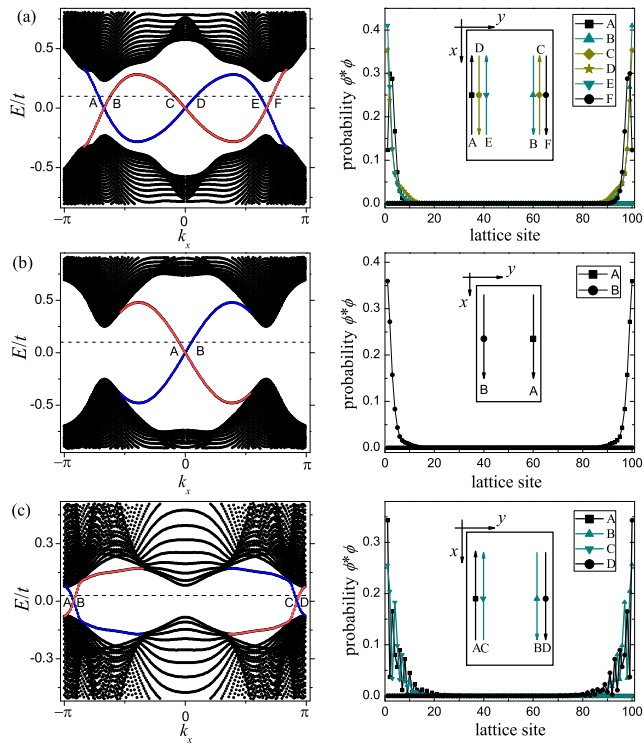}
\caption{\label{fig:5} (Color line) Quasiparticle spectra (left panels) and probability distributions of the edge states (right panels) for tFF[$-1$] (a), tFF[$1$] (b), and tFF[$-2$] (c) phases. Insets in the right panels show the directions of the currents corresponding to the edge states. From top to bottom row, the out-f-plane magnetic fields are $h_{z}/t=1.6$, $2$, and $3.1$, respectively. Other parameters are $t_{\rm so}/t=1$, $U/t=5$, $\mu/t=1$, and $h_{x}/t=0.1$.}
\end{figure}

To gain more insight into these topological phases, we further consider the edge modes of the system by imposing a hard-wall boundary condition along the $y$ direction. Fig.~\ref{fig:5} shows the quasiparticle spectra of distinct tFF phases and the probability distribution of the corresponding edge states. Consider, for example, the tFF[-1] phase in Fig.~\ref{fig:5}(a), the lines of the edge modes cross the Fermi level three times. To distinguish each edge state clearly, we plot a horizontal dashed line slightly above the Fermi level to reveal the Hall current. It shows that the dashed line intersects with the lines of edge modes at A, B, C, D, E, and F successively. Correspondingly, the probability distributions over the lattice sites for these states are plotted in the right panel of Fig.~\ref{fig:5}(a), which shows that states A, D, and E locate at the lower edge and states B, C, and F at the upper edge. These edge states also give rise to edge currents, $I\propto-\partial E/\partial k_{x}$, at the upper or lower edges. To proceed further, let us focus on the edge states A and F, which locate at upper and lower edges, respectively. Due to the translational symmetry along the $x$ direction, the band is symmetrical about the $k_{x}$ axis, which suggests that the group velocities of the states A and F have the same magnitude but with opposite signs. As a result, states A and F form a clockwise current loop at the edges and contribute a Chern number, $C_{\rm AF}=-1$, to the system. Similar analysis can be carried out for other pairs of states and leads to the Chern numbers $C_{\rm CD}=1$ and $C_{\rm BE}=-1$. Consequently, the total Chern number of the system is $C=C_{AF}+C_{CD}+C_{BE}=-1$. Likewise, from Fig.~\ref{fig:5}(b) and (c), it can be verified that the the Chern numbers of the tFF[1] and tFF[-2] phases are indeed $C=1$ and $-2$, respectively.

\section{Summary\label{sec:6}}
In summary, we have investigated a triangular lattice with SOC and in-plane
magnetic field. We construct the phase diagram in the noninteracting andinteracting
cases respectively. In the noninteracting case, we distinguish different phases by Chern numbers.
The Chern insulators with Chern number $C=1$ and $C=-2$ are limited within two ellipses.
In the interacting case, we get a very rich phase diagram including gapped and gapless topological FF superfluid
phases with Chern numbers $C=-1,1,-2$. The topological FF phases are characterized by phase modulating order
parameters and non-zero Chern numbers. We disclose edge states in different FF states and also discuss the connection between the edge state and the Chern number.

\section*{ACKNOWLEDGMENTS}
We acknowledge useful discussions with Yan He and Yuangang Deng. This work was supported by the NSFC (Grants No. 11074177, No. 11421063, and No. 11674334) and SRF for ROCS SEM (20111139-10-2).

\appendix
\section{Components of the BdG Hamiltonian}\label{appa}
The components of the BdG Hamiltonian (\ref{equ:3}) are
\begin{align}
a_{\mathbf{k}}&=2t(\cos{k_{1}}+\cos{k_{2}}\cos{Q}+\cos{k_{3}}\cos{Q})\text{,}\nonumber\\
b_{\mathbf{k}}&=2t(\sin{k_{2}}-\sin{k_{3}})\sin{Q}\text{,}\nonumber\\
c_{\mathbf{k}}&=\sqrt{3}t_{\rm so}(\sin{k_{3}}-\sin{k_{2}})\cos{Q}\text{,}\nonumber\\
d_{\mathbf{k}}&=\sqrt{3}t_{\rm so}(\cos{k_{2}}+\cos{k_{3}})\sin{Q}\text{,}\nonumber\\
e_{\mathbf{k}}&=t_{\rm so}(2\sin{k_{1}}-\sin{k_{2}}\cos{Q}-\sin{k_{3}}\cos{Q})\text{,}\nonumber\\
f_{\mathbf{k}}&=t_{\rm so}(\cos{k_{2}}-\cos{k_{3}})\sin{Q}\text{,}\nonumber
\end{align}
where $Q=\sqrt{3}Q_{y}/4$. It can be seen that $a_{\mathbf{k}}$, $d_{\mathbf{k}}$, and $e_{\mathbf{k}}$ are even functions of $\mathbf{k}$; while $b_{\mathbf{k}}$, $c_{\mathbf{k}}$, and $f_{\mathbf{k}}$ are odd functions.

\section{Roots of the equations $d_{x}(\mathbf{k})=d_{y}(\mathbf{k})=0$}\label{appb}
Explicitly, it can be easily evaluated that
\begin{align}
\mathbf{k}_{1}&=\left(0,\frac{2}{\sqrt{3}}\sin^{-1}\frac{\eta}{4}\right),\nonumber\\
\mathbf{k}_{2}&=\left(0,\frac{2}{\sqrt{3}}\left(\pi-\sin^{-1}\frac{\eta}{4}\right)\right),\nonumber\\
\mathbf{k}_{3}^{\pm}&=\left(\pm2\cos^{-1}{[-\frac{1}{2}\cos{(\frac{1}{2}\sin^{-1}{\eta}})]},\frac{1}{\sqrt{3}}\sin^{-1}{\eta}\right),\nonumber\\
\mathbf{k}_{4}^{\pm}&=\left(\pm2\cos^{-1}{[-\frac{1}{2}\cos{(\frac{1}{2}\sin^{-1}{\eta}})]},-\frac{1}{\sqrt{3}}\sin^{-1}{\eta}\right),\nonumber
\end{align}
where $\eta={2h_{x}}/{\sqrt{3}t_{\rm so}}$. The condition for existence of roots $\mathbf{k}_{1}$ and $\mathbf{k}_{2}$ is $\left|\eta\right|/4\leq1$ and $\left|\eta\right|\leq1$ for $\mathbf{k}_{3}^{\pm}$ and $\mathbf{k}_{4}^{\pm}$.


\begin{thebibliography}{99}

\bibitem{Cooper} Leon N. Cooper, Phys. Rev. \textbf{104}, 1189 (1956).

\bibitem{Fulde} P. Fulde and R. Ferrell, Phys. Rev. \textbf{135}, A550 (1964).

\bibitem{Larkin} A. I. Larkin and Yu. N. Ovchinnikov, Zh. Eksp. Teor. Fiz. \textbf{47}, 1136 (1964) [translation: Sov. Phys. JETP \textbf{20}, 762 (1965)].

\bibitem{Casalbuoni} R. Casalbuoni and G. Narduli, Rev. Mod. Phys. \textbf{76}, 263 (2004).

\bibitem{Kenzelmann} M. Kenzelmann, et al., Science 321, 1652 (2008).

\bibitem{Liao} Y.-A. Liao, A. S. C. Rittner, T. Paprotta, W. Li, G. B. Partridge, R. G. Hulet, S. K. Baur, and E. J. Mueller, Nature (London) \textbf{467}, 567 (2010).

\bibitem{Li} L. Li, C. Richter, J. Mannhart, and R. C. Ashoori, Nat. Phys. \textbf{7}, 762 (2011).

\bibitem{Hu1} H. Hu and X.-J. Liu, Phys. Rev. A \textbf{73}, 051603(R) (2006).

\bibitem{Radzihovsky}L. Radzihovsky and D. E. Sheehy, Rep. Prog. Phys. \textbf{73}, 076501 (2010).

\bibitem{Matsuda} Y. Matsuda and H. Shimahara, J. Phys. Soc. Jpn. \textbf{76}, 051005 (2007).

\bibitem{Shimahara} H. Shimahara, \emph{Theory of the Fulde-Ferrell-Larkin-Ovchinnikov State and Application to Quasi-Low-Dimensional Organic Superconductors}, in ed. by A.G. Lebed. \emph{The Physics of Organic Superconductors and Conductors}, Springer, Berlin (2008).

\bibitem{Koutroulakis}G. Koutroulakis, H. Kuhne, J. A. Schlueter, J. Wosnitza, and S. E. Brown, phys. Rev. Lett. \textbf{116}, 067003(2016).

\bibitem{Hasan} M. Z. Hasan and C. L. Kane, Rev. Mod. Phys. \textbf{82}, 3045 (2010).

\bibitem{Qi} X.-L. Qi and S.-C. Zhang, Rev. Mod. Phys. \textbf{83}, 1057 (2011).

\bibitem{Xiao} D. Xiao, M.-C. Chang, and Q. Niu, Rev. Mod. Phys. \textbf{82}, 1959 (2010).

\bibitem{Lin} Y.-J. Lin and K. Jimenez-Garcia, I. B. Spielman, Nature \textbf{471}, 83 (2011).

\bibitem{Wang} P. Wang, Z.-Q. Yu, Z. Fu, J. Miao, L. Huang, S. Chai, H. Zhai and J. Zhang, Phys. Rev. Lett. \textbf{109}, 095301 (2012).

\bibitem{Cheuk} L. W. Cheuk, A. T. Sommer, Z. Hadzibabic, T. Yefsah, W. S. Bakr and M.W. Zwierlein, Phys. Rev. Lett. \textbf{109}, 095302 (2012).

\bibitem{Wu} Z. Wu, L. Zhang, W. Sun, X.-T. Xu, B.-Z. Wang, S.-C. Ji, Y. Deng, S. Chen, X.-J. Liu, and J.-W. Pan, Nature \textbf{354}, 83 (2016)

\bibitem{Liu} X.-J. Liu and H. Hu, Phys. Rev. A \textbf{88}, 023622 (2013).

\bibitem{Qu} C. Qu, Z. Zheng, M. Gong, Y. Xu, L. Mao, X. Zou, G.-C. Guo, and C. Zhang, Nat. Commun. \textbf{4}, 2710 (2013).

\bibitem{Zhang} W. Zhang and Wei Yi, Nat. Commun. \textbf{4}, 2711 (2013).

\bibitem{Chen} C. Chen, Phys. Rev. Lett. \textbf{111}, 235302 (2013).

\bibitem{Cao} Y. Cao, S.-H. Zou, X.-J. Liu, S. Yi, G.-L. Long, and H. Hu, Phys. Rev. Lett. \textbf{113}, 115302 (2014).

\bibitem{Hu2} H. Hu, L. D, Y. Cao, H. Pu, and X.-J. Liu, Phys. Rev. A \textbf{90}, 033624 (2014).

\bibitem{Hatsugai} Y. Hatsugai, Phys. Rev. Lett. \textbf{71}, 3697 (1993).

\bibitem{Bernvig} B. A. Bernvig and T. L. Hughes, \emph{Topological Insulators and topological superconductors}, Princeton University, Princeton (2013).

\bibitem{Sarma} S. Yang, Z.-C. Gu, K. Sun, and S. Das Sarma. Phys. Rev. B \textbf{86}, 241112(R) (2012).

\bibitem{Huang} B. Huang, C. F. Chan, and M. Gong, Phys. Rev. B \textbf{91}, 134512 (2015).

%\bibitem{Kai1} K. Li, S.-L. Yu, and J.-X. Li, New J. Phys. 17 043032 (2015).

\bibitem{Kai2} K. Li, S.-L. Yu, Z.-L. Gu, and J.-X. Li, Phys. Rev. B \textbf{94}, 125120 (2016).

\bibitem{Iskin} M. Iskin, Phys. Rev. A \textbf{93}, 033632 (2016).

\bibitem{Xiongjun} X.-J. Liu, K. T. Law, and T. K. Ng, Phys. Rev. Lett \textbf{112}, 086401 (2014).

\bibitem{Zheng} Z. Zheng, M. Gong, X. Zou, C. Zhang, and G. Guo, Phys. Rev. A \textbf{87}. 031602(R) (2013).

\bibitem{momentqy} We have numerically verified that $Q_{x}$ is always $0$ in the parameter space covered in this work.

\bibitem{Altland} A. Altland and M. Zirnbauer, Phys. Rev. B \textbf{55}, 1142 (1997).

\bibitem{Schnyder} A.P. Schnyder, S. Ryu, and A. Furusaki, Phys. Rev. B \textbf{78}, 195125 (2008).

\bibitem{Ryu} S. Ryu, A. P. Schnyder, A. Furusaki, and A. W. W. Ludwig, New J. Phys. \textbf{12}, 065010 (2010).

\bibitem{Kiteav1} A. Kitaev, AIP Conf. Proc. \textbf{1134}, 22 (2009).

\bibitem{Thouless} D. J. Thouless, M. Kohmoto, M. P. Nightingale, and M. den Nijs, Phys. Rev. Lett. \textbf{49}, 405 (1982).

\bibitem{Tretiakov} O. A. Tretiakov and O. Tchernyshyov, Phys. Rev. B \textbf{75}, 012408 (2007).

\bibitem{Clement} C. H. Wong and R. A. Duine, Phys. Rev. A \textbf{88}, 053631 (2013).

\bibitem{Xu} Y. Xu, C. Qu, M. Gong, and C. Zhang, Phys. Rev. A \textbf{89}, 013607 (2014).

\bibitem{Fan} F. Wu, G.-C. Guo, W. Zhang, and W. Yi, Phys. Rev. Lett. \textbf{110}, 110401 (2013).

\bibitem{Deng} Y. Deng, T. shi, H. Hu, L. You, and S. Yi, arXiv:1607.05109

\bibitem{Kiteav2} A. Yu. Kitaev, Phys.-Usp. \textbf{44}, 131 (2001).

\bibitem{Ghosh} P. Ghosh, J. D. Sau, S. Tewari, and S. Das Sarma, Phys. Rev. B \textbf{82}, 184525 (2010).



\end{thebibliography}
\end{document}